\documentclass{aastex631}
\usepackage{cleveref} 
\usepackage{appendix}
\usepackage{graphicx}

\begin{document}

\title{Using machine learning method for variable star classification using the TESS Sectors 1-57 data}

\correspondingauthor{Kai Li}
\email{kaili@sdu.edu.cn}

\author{Li-Heng Wang}
\affiliation{Shandong Key Laboratory of Optical Astronomy and Solar-Terrestrial Environment, School of Space Science and Technology, Institute of Space Sciences, Shandong University, Weihai, Shandong 264209, China}

\author{Kai Li}
\affiliation{Shandong Key Laboratory of Optical Astronomy and Solar-Terrestrial Environment, School of Space Science and Technology, Institute of Space Sciences, Shandong University, Weihai, Shandong 264209, China}

\author{Xiang Gao}
\affiliation{Shandong Key Laboratory of Optical Astronomy and Solar-Terrestrial Environment, School of Space Science and Technology, Institute of Space Sciences, Shandong University, Weihai, Shandong 264209, China}

\author{Ya-Ni Guo}
\affiliation{Shandong Key Laboratory of Optical Astronomy and Solar-Terrestrial Environment, School of Space Science and Technology, Institute of Space Sciences, Shandong University, Weihai, Shandong 264209, China}

\author{Guo-You Sun}
\affiliation{Xingming Observatory, Urumqi, Xinjiang, China }



\begin{abstract}

The Transiting Exoplanet Survey Satellite (TESS) is a wide-field all-sky survey mission designed to detect Earth-sized exoplanets.
After over four years photometric surveys, data from sectors 1-57, including approximately 1,050,000 light curves with a 2-minute cadence, were collected. By cross-matching the data with Gaia's variable star catalogue, we obtained labeled datasets for further analysis. Using a random forest classifier, we performed classification of variable stars and designed distinct classification processes for each subclass, 6770 EA, 2971 EW, 980 CEP, 8347 DSCT, 457 RRab, 404 RRc and 12348 ROT were identified. Each variable star was visually inspected to ensure the reliability and accuracy of the compiled catalog. Subsequently, we ultimately obtained 6046 EA, 3859 EW, 2058 CEP, 8434 DSCT, 482 RRab, 416 RRc, and 9694 ROT, and a total of 14092 new variable stars were discovered.

\end{abstract}

\keywords{stars:variable; surveys; catalogues; methods: data analysis}


\section{Introduction} \label{sec:intro}

Variable stars refer to a type of celestial body that exhibits periodic changes in luminosity due to reasons such as occultation, pulsation, or rotation. Variable stars, characterized by their distinctive variability in luminosity, offer astronomers a valuable tool for understanding the internal structures of stars, their evolutionary processes, and fundamental stellar physics theories. Therefore, researchers aim to identify as many variables as possible to assist the study of stellar evolution, exploring galactic structure, and so on. As early as hundreds of years ago, people began to explore variables. In recent decades, advancements in CCD technology and large-scale sky surveys have led to a significant increase in the discovery of variable stars. Using large-scale ground photometric observation equipment, such as: the Opti-cal Gravitational Lensing Experiment \cite[OGLE;][]{r1}, Vista Variables in the Via Lactea \cite[VVV;][]{r3}, the All-Sky Automated Survey for Supernovae \cite[ASAS-SN;][]{r2}, the Asteroid Terrestrial-impact Last Alert System \cite[ATLAS;][]{atlas1, atlas2} , Super Wide Angle Search for Planets \cite[SuperWASP;][]{wasp} and the Zwicky Transient Facility (ZTF) survey \citep{ztf1,ztf2},  thousands of variable stars have been discovered. However, the accuracy of the ground-based observations is often affected by the weather. Space-based observations, such as Kepler space telescope \citep{r4}, Gaia \citep{r5} and The Transiting Exoplanet Survey Satellite \cite[TESS;][]{s1}, have alleviated constraints related to observing locations, weather and atmosphere, obtained a vast amount of data. Faced with the large amount of data, manual classification of variables becomes impractical, demanding computational assistance to enable automated classification. In this context, machine learning emerges as an effective tool for classifying light curves.

A variety of methods have been proposed to carry out this work, such as Random Forest \cite[RF;][]{r6} and deep learning \citep{r7} in supervised learning, Density-Based Spatial Clustering of Applications with Noise (DBSCAN) and K-means in unsupervised learning, as well as semi-supervised methods. All of these approaches have demonstrated excellent performance in this study.
While black-box machine learning methods can achieve better classification results, they often come at the cost of significantly compromising physical interpretability—a trade-off that researchers aim to avoid. Correspondingly, the RF algorithm requires artificially designed features as input for training, such as Fourier decomposition, skewness, average, color, etc. This implies that, in comparison to deep learning, more effort is required to extract features and demonstrate their effectiveness. 

The exponential growth of photometric survey data has driven researchers to increasingly adopt machine learning techniques for variable star classification.  For example, \cite{r2} has completed the variable star classification of ASAS-SN, and \cite{r8} uses the DBSCAN method to achieve the variable star classification of ZTF data. In recent years, a series of variable star classification studies have been carried out using TESS data, such as, \cite{EB_4000} completed the classification of TESS eclipsing binaries, \cite{tess12000} carried out variable star classification through visual inspection, and \citep{tess_p} obtained short-period pulsating hot-subdwarf stars. However, there is still a lack of a comprehensive classification that covers all types of variable stars using TESS data. Therefore, it is necessary to conduct the work of classifying variable stars using TESS data. 

In this paper, we used the TESS 2-min data with the goal of completing the classification of variable stars. Through interpretable features and the physical properties. This article is constructed as follows: in Section 2, we provide an overview of the TESS 2-min data, including relevant information and data characteristics, and perform period calculation and feature extraction. Section 3 details our classification process. In Section 4, we present and analyze our classification results. Finally, in Section 5, we summarize this work.

\section{ DATA AND METHODS} \label{sec:dam}
\subsection{Data }\label{Data}

TESS \citep{s1} is an all-sky survey satellite mainly used to search for planets similar to the size of the Earth. The TESS survey conducts sky observations in an area of 24° × 96°. The observation time of each sector is approximately 27.4 days. In addition to the observation data at a 2-minute cadence, TESS will acquire a full-frame image every 30 minutes. In the first year of the mission, TESS observed the southern ecliptic hemisphere, including sectors 1-13; in the second year, it observed the northern ecliptic hemisphere, including sectors 14-26; in the third year, it re-observed the southern ecliptic hemisphere, including sectors 27-39; in the fourth year, part of the northern ecliptic hemisphere was re-observed, and the 240° ecliptic strip was observed for the first time. This observation included 16 sectors, namely sectors 40-55; and the rest of this article Sectors 56-57 are the fifth year of continued observation of the northern ecliptic hemisphere. We downloaded the TESS 2-min light curve data from the MAST website\footnote{MAST: \url{https://archive.stsci.edu/tess/bulk_downloads/bulk_downloads_ffi-tp-lc-dv.html}} \citep{https://doi.org/10.17909/t9-nmc8-f686} and extracted the $PDCSAP\_FLUX$ (systematically corrected flux data) from the fits files. We obtained approximately 1,050,000 light curves, and saved them by sectors.

In this work, we focus on 3 different main variable star types: eclipsing binaries \cite[$EB_s$,][]{ebs_1,ebs_2,ebs_3},  pulsations \citep{cep_1,cep_2} and Rotational variables (ROT). The main variable star classes are divided into the sub-classes. According to filling factor of the Roche lobe, eclipsing binaries can be classified into types detached (EA), semidetached (EB), and contact (EW). Owing to the challenges in delineating clear boundaries among these three categories, we classified eclipsing binaries into two types: EA and EW. And We classify pulsations into: Cepheid (CEP), $\delta$ Scutis (DSCT), ab-type RR Lyrae (RRab) and c-type RR Lyrae (RRc).

To acquire the necessary labeled data for our machine learning models, we performed a cross-match between our dataset and the Gaia DR3 variable star catalog \citep{q1}, employing a matching radius of 3 arcseconds.  Since we only require the main type for our labeled data, this star catalog meets our needs effectively. After the cross-match, we obtained a total of  9022 $EB_s$, 1528 pulsations and 3744 ROT. It should be noted that, due to overlapping observation sectors, these data contain duplicates.  Rather than removing duplicates, we used them as input for separate targets due to the variations in observation frequency and noise levels across different sectors. Thus, they serve as valuable "data augmentation" for training.

\subsection{Period Determination }\label{period d}
We use the Generalized Lombe-Scargle \cite[GLS;][]{s2,s3} method to perform periodic searches. The GLS algorithm from Astropy \citep{s4, asp1_0} was utilized to achieve this. We completed the period search in two steps: first, we performed a period search within the range of 0.01 to 33.33 days, setting the nterms value to 1 to control the number of Fourier terms used in the model. Each TESS sector spans approximately 27 days; thus, setting the maximum search duration to 33 days ensures coverage of a complete sector. Although our dataset includes multiple observations from different sectors, we opted to use the light curve from one sector to maintain data consistency and simplify processing. Subsequently, we performed a second period search near the frequency corresponding to the maximum power identified in the initial search by setting nterms to 5 to achieve higher precision.  We narrowed the frequency range to less than 2 and used 5000 grids for the period search. This approach enhances the precision and reliability of the period determination. If the initial frequency $f_{0}$ is less than 1, we will search in the range [$f_{0}$/1.9,  $f_{0}$×2].

Using this method, we obtained the period for each star. Additionally, we recorded the maximum power values from the power spectrum obtained in the first and second calculations as $power_1$ and $power_2$. This allows us to assess to some extent whether the data exhibits periodicity, the closer the $power$ is close to 1, the greater the likelihood that the data is periodic. However, We found that, although this processing method can distinguish variables from non-variables to a certain degree, it still results in the loss of variables. Since our goal is to identify as many variables as possible, we didn't try to separate variables from non-variables at this stage. Instead, we saved the data for further analysis. It is worth mentioning that for eclipsing binaries, the period returned by the GLS period diagram is often 0.5 times the true period, so we directly doubled the period during the subsequent phase folding and Fourier parameters calculations.

Considering that TESS is a systematics-limited instrument, periodograms often show peaks around periods of 7 days as it is half an orbit.  Therefore, we randomly selected some typical variable stars and examined their GLS periodograms, as shown in \Cref{fig:LS_p}  (two targets as examples). We did not detect any significant periodic signals at 7 days, and have therefore ruled out this possibility. 

We performed a statistical analysis on the labeled data. The period distribution for each category is illustrated in \Cref{fig:p_dis}. Furthermore, we calculated the signal-to-noise ratio (SNR) of GLS periodograms for the labeled data, and the distribution of SNR  is presented in \Cref{fig:snr}.

\subsection{Other features }\label{other features}
Feature extraction plays a very important role in the entire classification process. A well-chosen set of features can significantly enhance the classifier's accuracy. However, poorly selected features can lead to overfitting, which reduces both the classifier's performance and its ability to generalize. In addition to the period, we also introduce related parameters such as phase difference, amplitude ratio, skew, etc., to aid in the classification process.

First, we performed flux normalization using the median flux as a reference. Next, we conducted phase folding by establishing the minimum flux at phase 0. We did not restrict the data points to the 1\%-99\% range, as described in some papers, because we found that this approach would remove part of the eclipse data for EA-type $EB_s$, which will introduce unnecessary errors. To obtain the Fourier parameters, we use the following equation to fit the light curve with a Fourier series:
\begin{equation}
    \hat{y} = a_0 + \sum_{i=1}^{4}a_i sin(2\pi f) + b_i cos(2\pi f),
\end{equation}
where $a_0$ represents the constant term, which is the average value of the data; $a_i$ represents the sine component of the i-th item; $b_i$ represents the cosine component of the i-th item, and $f$  represents the frequency of the variable star, since we transferred time to phase, so the frequency here should be 1.

While we defined the minimum flux point as phase 0, systematic photometric uncertainties and instrumental response variations may introduce phase-alignment discrepancies.  As a result, differences in the observation starting points can lead to variations in the Fourier coefficients. After completing the fitting, we convert it into the form of amplitude and phase through the following equation:
\begin{equation}
    A_i = \sqrt{a_i^{2} + {b_i^{2}} }
\end{equation}
\begin{equation}
    \phi_i = arctan(b_i/a_i)
\end{equation}
where $A_i$ represents the amplitude of the i-th item, and $\phi_i$ represents the phase of the i-th item.

We followed the method described by \cite{r8} and calculated the amplitude ratio and phase difference $A_{ij} = A_i/A_j$, $\phi_{ij} = \phi_i-\phi_j*i/j$.
To avoid the impact of erroneous data, we didn't choose to directly use the difference between the maximum and minimum values of the raw data when calculating the amplitude. We determined the amplitude via a fourth-order Fourier fit. After the fitting is completed, we obtain the parameter of goodness of fit $r^2$, which can reflect the quality of the fitting to a certain extent. In addition, we also added skew, kurtosis and average, which can help us analyze the data statistically.

We aim to introduce additional information, such as temperature, magnitude, color, etc., that have been proven to be useful for classification in other studies. However, due to the large amount of missing data, these additional parameters were ultimately not adopted. 

\section{ Classification } \label{thrid:Class}

The use of RF to complete classification tasks has proven to be very effective. 
For instance, the RF algorithm has been effectively utilized to classify light curves in previous studies, such as \cite{t1}, \cite{r2}. 
The generalization and robustness of RF are widely recognized, offering high accuracy. Additionally, compared to neural network architectures such as Convolutional Neural Networks (CNNs) and Recurrent Neural Networks (RNNs), RF provides faster training speed and can achieve high accuracy in a shorter time. Therefore, we used RF model for the classification task. Although using subclasses directly for classification resulted poor performance, especially in terms of accuracy, we improved the situation by addressing the imbalance in the data. Specifically, we reduced the overall size of the training set to mitigate the issue. This reduction will make it challenging for the model to achieve satisfactory training results.  Therefore, we first use RF to classify the data into four main categories: $EB_s$, pulsations, ROT, and non-variables, and then we performed sub-class classification for each category. 

\subsection{Random Forest Classification}\label{random}
RF is an ensemble machine learning method that combines multiple classifiers to form a more effective ensemble classifier. It primarily uses the bagging (Bootstrap Aggregating) technique. In this approach, $n$ training samples are drawn with replacement from the original training set, and these samples are used to train a weak classifier. By repeating this process multiple times, an ensemble of classifiers is created. Each classification result is determined by a majority vote among these multiple classifiers. In the context of RF, the weak classifier is typically a decision tree. Compared to other classifiers, RF not only achieves very high accuracy on large datasets but also handles high-dimensional feature directly, thereby avoiding the information loss that can occur with dimensionality reduction. 

We employ the Random Forest algorithm from scikit-learn \citep{t2} to develop a classifier for the four aforementioned categories. Our training samples are as described in Section \ref{Data}, and the label data of variable stars are obtained through cross-matching with Gaia variable star catalog. For non-variable star data, we set an indicator: $power_2<$0.01, $r^2<$0.01, we consider stars inside the range of $power_2$ and $r^2$ to be non-variable stars. Since we used this method to obtain non-variable stars, we can no longer use the $power$ and $r^2$ parameter as training for RF. To ensure balanced training data, we set the number of training samples for each of the four types of labels to 1,000, resulting in a total of 4,000 training curves. 
Given the presence of classification errors in Gaia, we aimed to remove potentially misclassified variable stars prior to classification. 
Considering that the pulsating variables can obtain a higher goodness of fit when performing Fourier fitting, we limit the value of $r^2$ to 0.64 to ensure that at least 1000 data can be used for model training. Due to the aforementioned restrictions potentially leading to the removal of a large number of EA type EBs and ROT with low signal-to-noise ratios, no restrictions are applied to $EB_s$ and ROT. 

We divide our data set into 80\% for training and 20\% for testing, setting the RF parameters $n\_estimators=700$, $max\_features=5$. To evaluate the reliability of our training model, we use Out-Of-Bag (OOB) evaluation. This method involves using out-of-bag samples—those not included in the training subset during each sampling—for prediction with the already trained decision trees. To determine the OOB score, we compute the accuracy of each decision tree and the average values. This score aids in identifying the optimal model and offers a visual comparison to assess performance effectively. Besides, we also provide common evaluation indicators for results: precision and recall.

\begin{figure}
    \centering
    \includegraphics[width=0.5\linewidth]{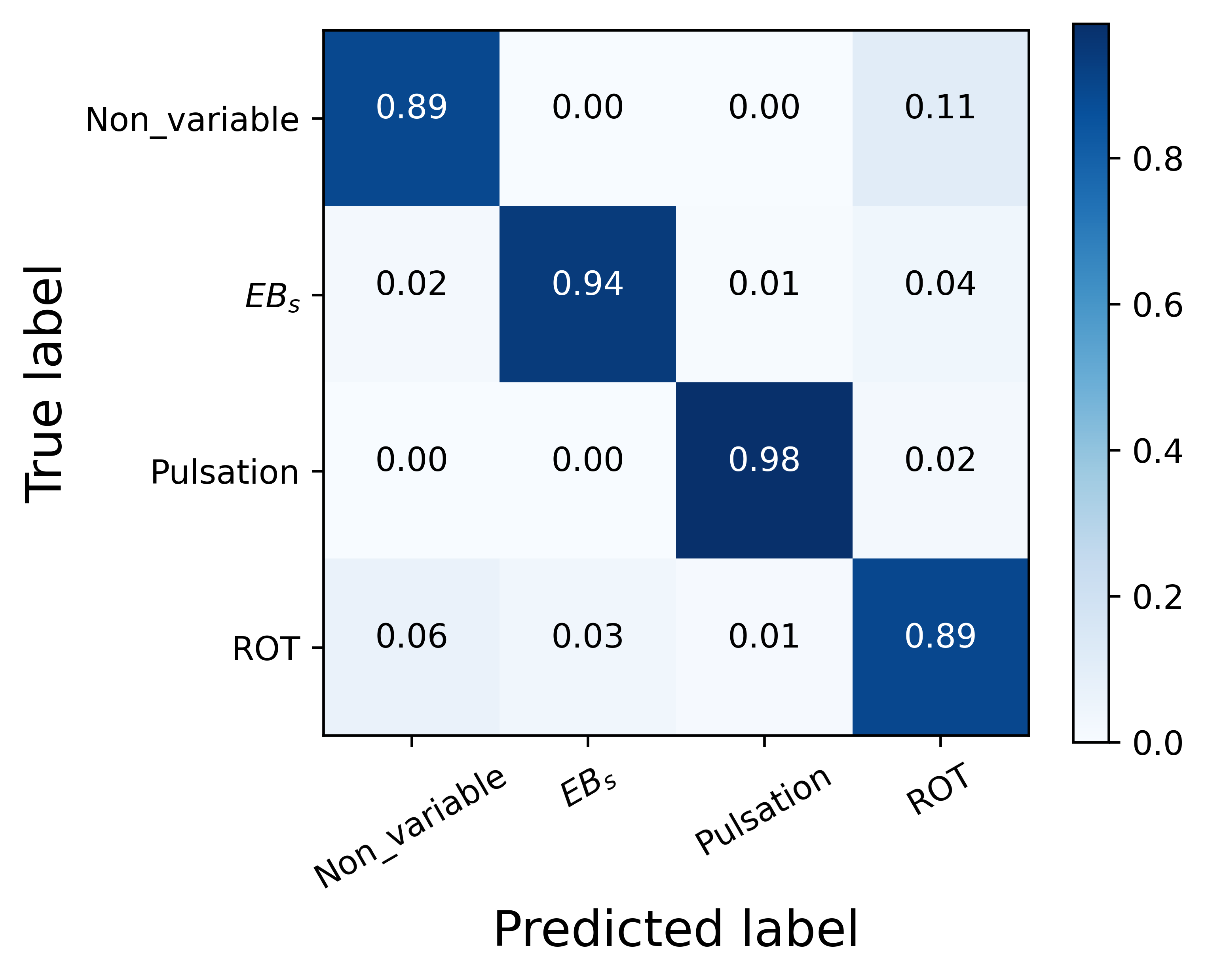}
    \caption{Confusion matrix for training a RF classifier with labeled data, with the x-axis being the predicted category and the y-axis being the input category}
    \label{fig:fig1}
\end{figure}

In the end,  we obtained a model with  $oob\_score$ at $0.9178$. Then we used the model to predict the test set. The test set has a total of 800 targets,  we provide a confusion matrix as shown in \Cref{fig:fig1},
the precision and recall are shown in \Cref{tab:label1}.

\begin{figure}
    \centering
    \includegraphics[width=0.5\linewidth]{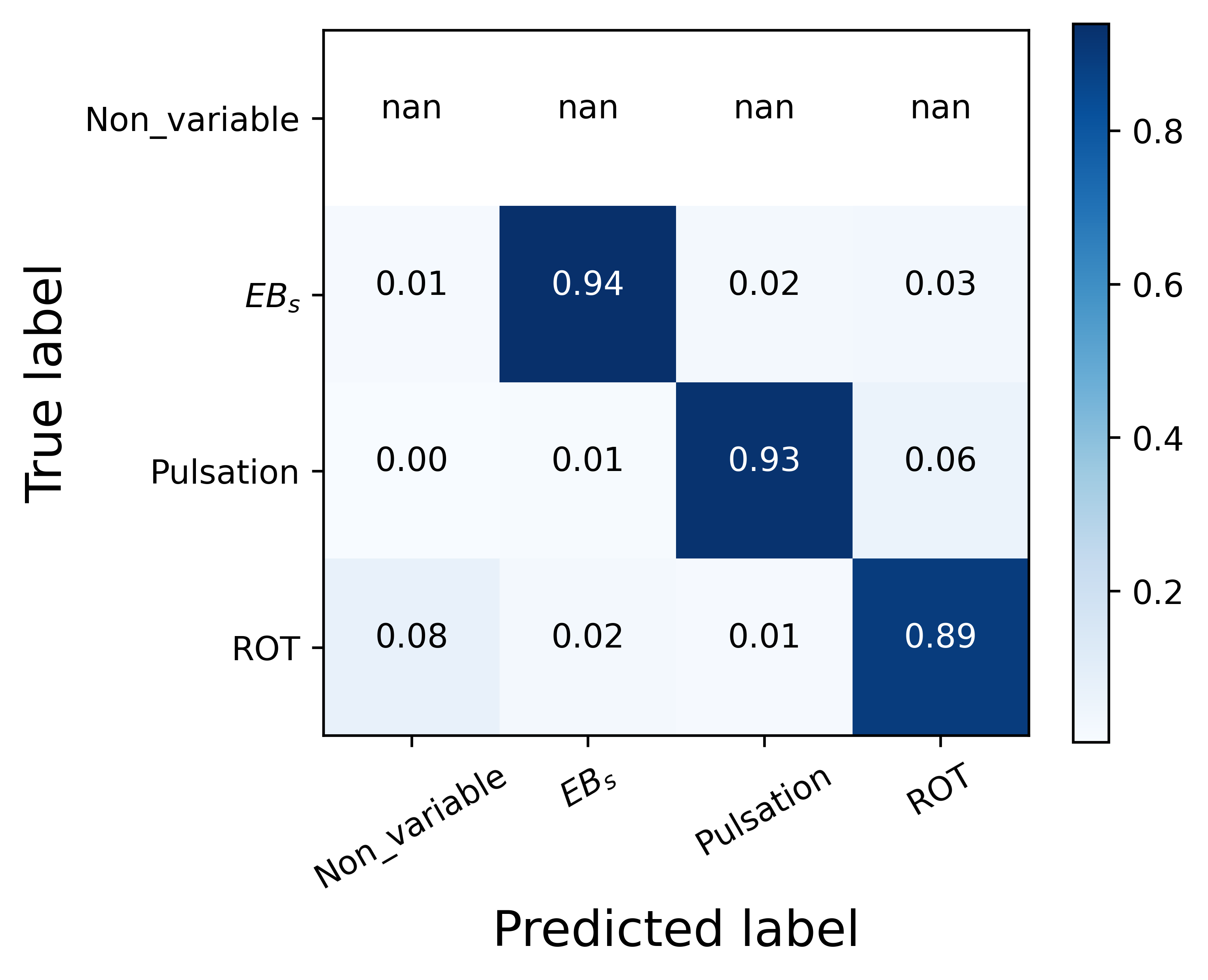}
    \caption{The confusion matrix we got after taking all labeled data (except non-variable stars) as input.}
    \label{fig:fig2}
\end{figure}

To further evaluate our model's performance, we inputted all labeled light curves into model for prediction. Notably, non-variable stars were excluded from this analysis since they are not available in the Gaia variable star catalog. In total, we used 14,294 light curves for prediction. The specific numbers of each type are as described in Section \ref{Data},  the confusion matrix obtained is shown in \Cref{fig:fig2} .

\begin{table}[h]
\caption{Model precision and recall}
\centering
\begin{tabular}{ccc}
\hline
class & precision & recall \\
\hline
$non\_variable$&92\%&89\%\\
$EB_{s}$& 96\%& 94\%\\
$Pulsation$& 98\%& 98\%\\
$ROT$&85\%&89\%\\
\hline
\end{tabular}
\label{tab:label1}
\end{table}


We used the established RF model to predict all the data, and finally obtained approximately 580,000 non-variable stars, about 50,000 $EB_s$, 27,000 pulsations, and 450,000 ROT. Considering that our training data are labels provided by Gaia, and there are certain differences between the two surveys. A visual inspection revealed that numerous non-variable stars, especially within the ROT category, are interspersed among the classifications of variable stars. This indicates potential discrepancies between the actual dataset and the labeled dataset. Additionally, since the classifier can generate errors, further refinement and classification of the data will be necessary in next work. Then, we also obtained the likelihood scores provided by the classifier for each category.

\subsection{$EB_s$}\label{EB}

We present a more detailed subclassification of $EB_s$ in this section. In the past studies, eclipsing binary stars were divided into different subcategories: for example, \cite{r8} divided $EB_s$ into two types: EA and EW, and \cite{t3} divided them into three types: EA, EW, and ELL. However, due to the limited information provided by the light curves, EA and EB or EW have been confused. Therefore, following \cite{r8}, we categorize $EB_s$ into two types: EA and EW.

\begin{figure}
    \centering
    \includegraphics[width=0.5\linewidth]{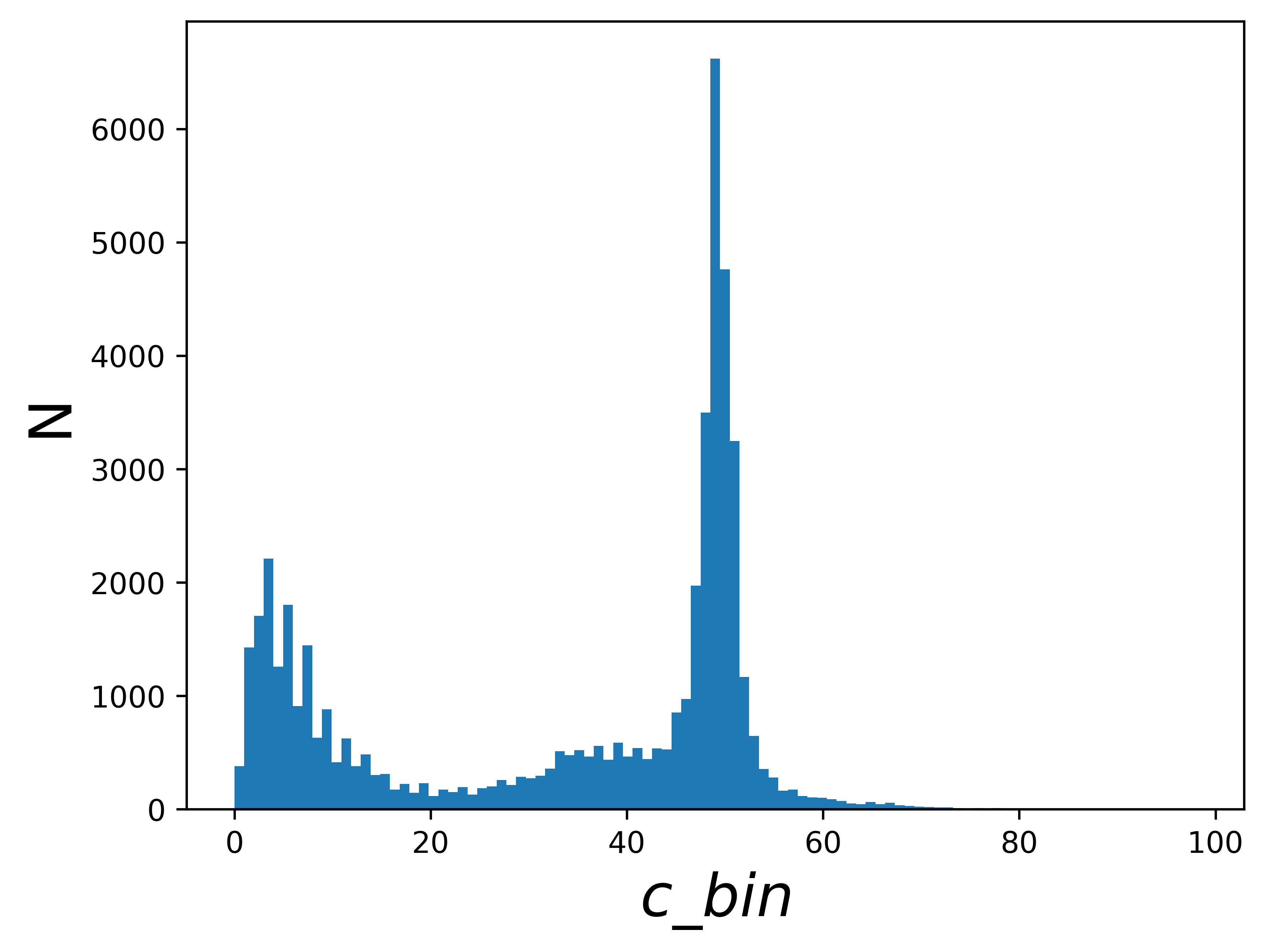}
    \caption{Statistical chart of $c\_bin$ parameters (normalize the light curves and then count the number of points with values below 0.5). The abscissa is the parameter value and the ordinate is the statistical number.}
    \label{fig:fig3}
\end{figure}

EA type $EB_s$ exhibit deeper eclipses, whereas the light curves of EW type $EB_s$ have a more sinusoidal shape, and a statistical feature $c\_bin$ is designed here: firstly, bin the light curves to 100 points, then scale them to the 0-1 interval, and then count the number of data points below 0.5, this value is called $c\_bin$. According to the original concept, if the light curve undergoes sinusoidal variation or follows a uniform distribution (with each data point having an equal chance of being positioned anywhere within the distribution), such a statistical measure is expected to vary near 50. In addition, due to the relatively short duration of eclipses within one period for EA type $EB_s$, the statistical value will be lower. In order to find a suitable threshold, the $c\_bin$ parameters of each star are counted here. The statistical chart is shown in \Cref{fig:fig3}. It can be seen that there is an extremum at 20, so $c\_bin=20$ was chosen to divide the data. We classify the data with $c\_bin<=20$ as EA, and the data with $c\_bin>20$ as EW. The $EB_s$ obtained by this method will be mixed with non-variable stars (although a large number of non-variable stars are removed in Section \ref{random}, it is still difficult to avoid partial aliasing), thus the $r^2$ parameter is used for elimination. For Fourier fitting, the $r^2$ of non-variables will be closer to 0. Consequently, only those data with an $r^2$ below 0.1 are excluded from consideration.  Finally, the classification of EA and EW was obtained. Interestingly, a large number of variables with periods shorter than 0.12 days have been found in EW. According to the study of  \cite{t4}, the period of the EW type $EB_s$ should not be shorter than 0.12 days. This phenomenon is likely due to the very similar light curve shapes of EW and DSCT, causing stars with periods shorter than 0.12 days to be misclassified as EW rather than DSCT. Finally, after removing duplicate data, we obtained 6770 EA and 2971 EW.

\subsection{Pulsations}\label{Pulsation}

Pulsations are stars whose light curves exhibit periodic changes due to pulsations. A key characteristic of these stars is the period-luminosity relationship: the period is proportional to their luminosity, which serves as an important feature for distinguishing pulsating variables. However, due to the lack of absolute magnitude for some targets, we need to use other methods to complete the classification. Considering that the observation of each sector spans approximately 27 days, long-period variable stars cannot be detected.  Based on the classifications of pulsations described in most works \citep[e.g.,][]{r2,r8}, we divide pulsations into four categories: CEP, DSCT, RRab, and RRc.  The period of DSCT is approximately less than 0.3 d, the period of RRab is approximately between 0.42-1 d, and the period of RRc is approximately between 0.2-0.42 d. Since the previous three categories are all short-period types, we classied all pulsations with periods longer than 1 d as CEP types. 

\begin{figure}
    \centering
    \includegraphics[width=1\linewidth]{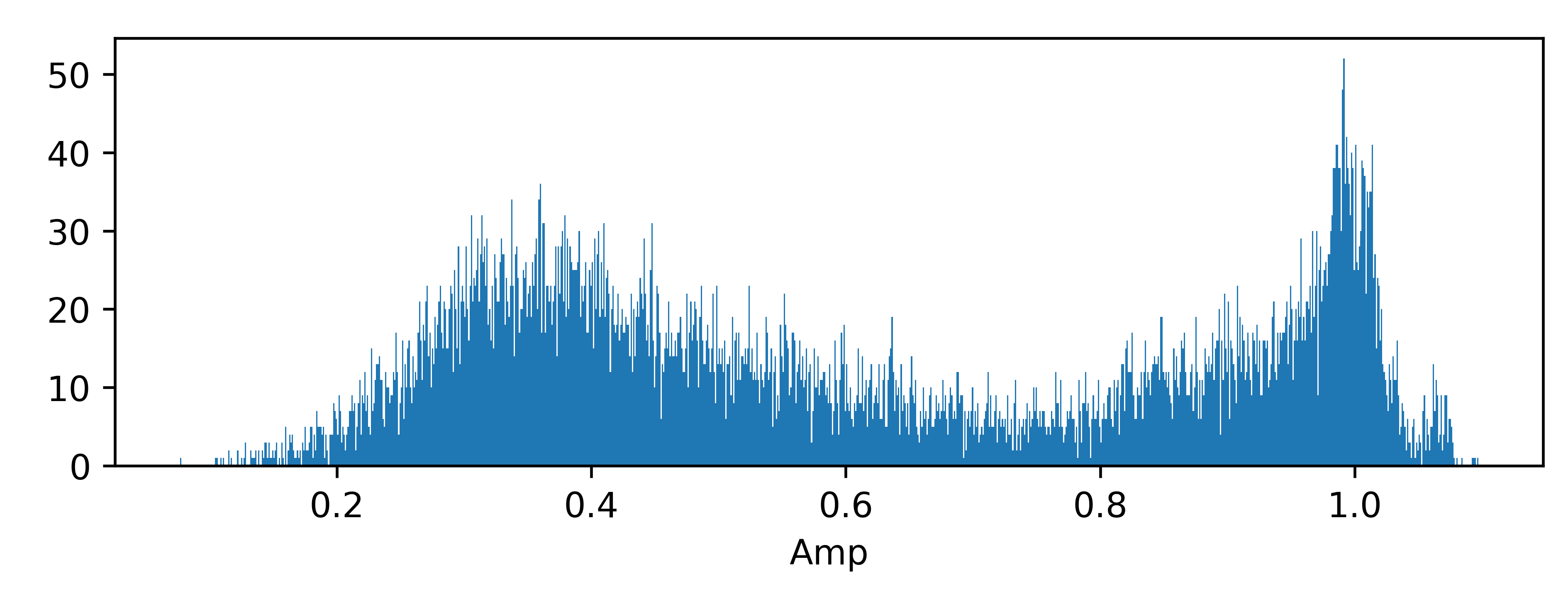}
    \caption{This figure is the statistical distribution diagram of amplitude.}
    \label{fig:fig4}
\end{figure}

The elimination of non-variable stars is also a crucial task. 
We used the normalization method mentioned in Section \ref{other features} to standardize the amplitude of each light curve. 
In order to distinguish variables and non-variables, following the method mentioned in Section \ref{EB}, the light curve is first folded and binned to 100 points, and then scaled to the 0-1 interval. And we performed fourth-order Fourier fitting. The statistics of amplitude is shown in \Cref{fig:fig4}. Theoretically, for variables, the amplitude should be close to 1; for non-variable stars, the amplitude should be smaller. It appears that selecting an appropriate threshold is more straightforward based on the amplitude. 
Consequently, a 100th-order polynomial fitting was applied to the amplitude statistics to determine  the minimum value of the fitted curve. The desired minimum value was then visually selected, resulting in an amplitude threshold being set to 0.75. 

Although we aimed to include multiple subclasses in the training dataset, the availability of labeled data was limited to RR Lyrae and CEP due to cross-matching with Gaia's variable star catalog. Consequently, our training data exclusively comprised these two classifications. We used 150 samples of each type, totaling 300 samples for training. The dataset was divided into an 80\% training set and a 20\% test set.  Set $n\_estimators=100$, $max\_features=5$. We use the relevant parameters obtained from the above Fourier decomposition for training, and the final $oob\_score=0.9833$ is obtained. We found that the vast majority of variables with periods less than 0.3 d were classified as RR Lyrae, indicating that most DSCT were misclassified as RR Lyrae. Therefore, further classification is needed for this type of variable. 

A dataset consisting of DSCT, RRab, and RRc types has been compiled. To distinguish DSCT from RR Lyrae, a small RF model was trained. Due to the distinct characteristics of RRab and RRc, they were trained as separate categories. In order to obtain the initial training samples, these three types of unique period intervals were selected as much as possible: the period of DSCT is less than 0.14 days, the period of RRc is [0.32, 0.37] days, and the period of RRab is [0.45, 1] days. We gathered 200 variable stars for each category, totaling 600 samples. Since the period is used to construct the dataset, they are no longer suitable for model training. Therefore, only Fourier correlation parameters were used for training. The training set and test set account for 80\% and 20\% respectively. Set $n\_estimators=200$, $max\_features=4$, and finally get $oob\_score=0.76875$. The results were suboptimal, likely due to the insufficiency of Fourier decomposition features alone for effective differentiation. Therefore, to improve the classification accuracy of DSCT, we introduced a period constraint based on the conclusions from \cite{gaia_dr2} and \cite{r2}: targets labeled as DSCT with periods greater than 0.3 days were excluded, while targets labeled as other types with periods less than 0.2 days were included. 

Finally, we classified RRab and RRc based on the shape of their light curves. RRab stars typically exhibit asymmetric light curves, while RRc stars display nearly sinusoidal ones. To determine the differences between them, we fitted a single sinusoidal curve to the binned data.  Theoretically, RRc should yield a higher goodness of fit. We calculated the goodness of fit for the light curves, fitted the resulting statistical data with a 50th-order polynomial, and identified the minimum value of fitted curve. We set the threshold for classification at 0.868 to differentiate between RRab and RRc. 

Meanwhile, we examined the distribution of variable star types and quantities with periods under 0.12 days across the ASAS-SN and ZTF catalogs.  In the ASAS-SN catalog \citep{r2}, there are 5833 targets with periods less than 0.12d, of which 5729 are classified as DSCT. In the ZTF catalog \citep{r8}, there are 15316 targets with periods less than 0.12d, of which 14143 are classified as DSCT. Therefore, we consider all targets with periods shorter than 0.12d, except for EA, to be DSCT. 
In this part of the work, a total of 980 CEP, 8347 DSCT (including ROT with periods shorter than 0.12 days), 457 RRab, and 404 RRc were obtained.

\subsection{ROT}\label{Rotation}

ROT variable stars exhibit periodic luminosity changes due to variations in surface temperature caused by their rotation. The surface magnetic activity can occur at any location and time, leading to a diverse shapes of light curves. As these light curves often closely resemble those of other variable star types, accurate classification based solely on morphology is challenging. Consequently, rather than attempting to obtain the different sub-classes within the ROT category, we focus on identifying and removing non-variable stars from this group. Due to the near-sinusoidal shape and clear periodicity of ROT light curves, both the goodness of fit from Fourier fitting and the power value are numerically close to 1.  Therefore, $power_2>0.7$ and $r^2>0.7$ are directly used as the filtering conditions for the real ROT. As we mentioned in the previous subsection, targets with periods shorter than 0.12 days are considered as DSCT.
Finally, a total of  12, 348 ROT were obtained.

\section{Classification results} \label{sec:result}
\begin{figure}
    \centering
    \includegraphics[width=1\linewidth]{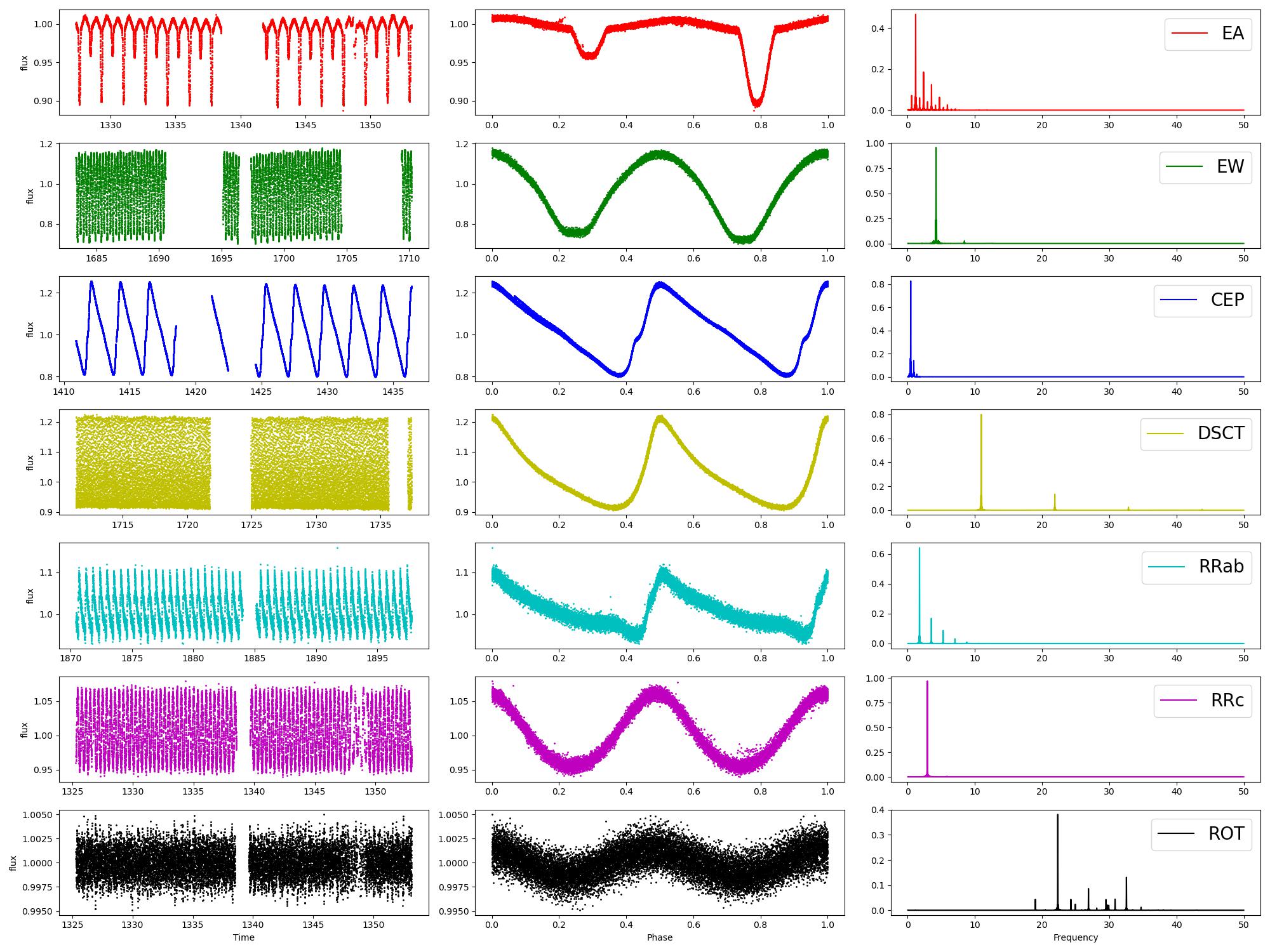}
    \caption{The typical light curve for each category, from left to right: the original light curve, phase folded curve, and GLS periodogram.}
    \label{fig:light}
\end{figure}

\begin{figure}
    \centering
    \includegraphics[width=1\linewidth]{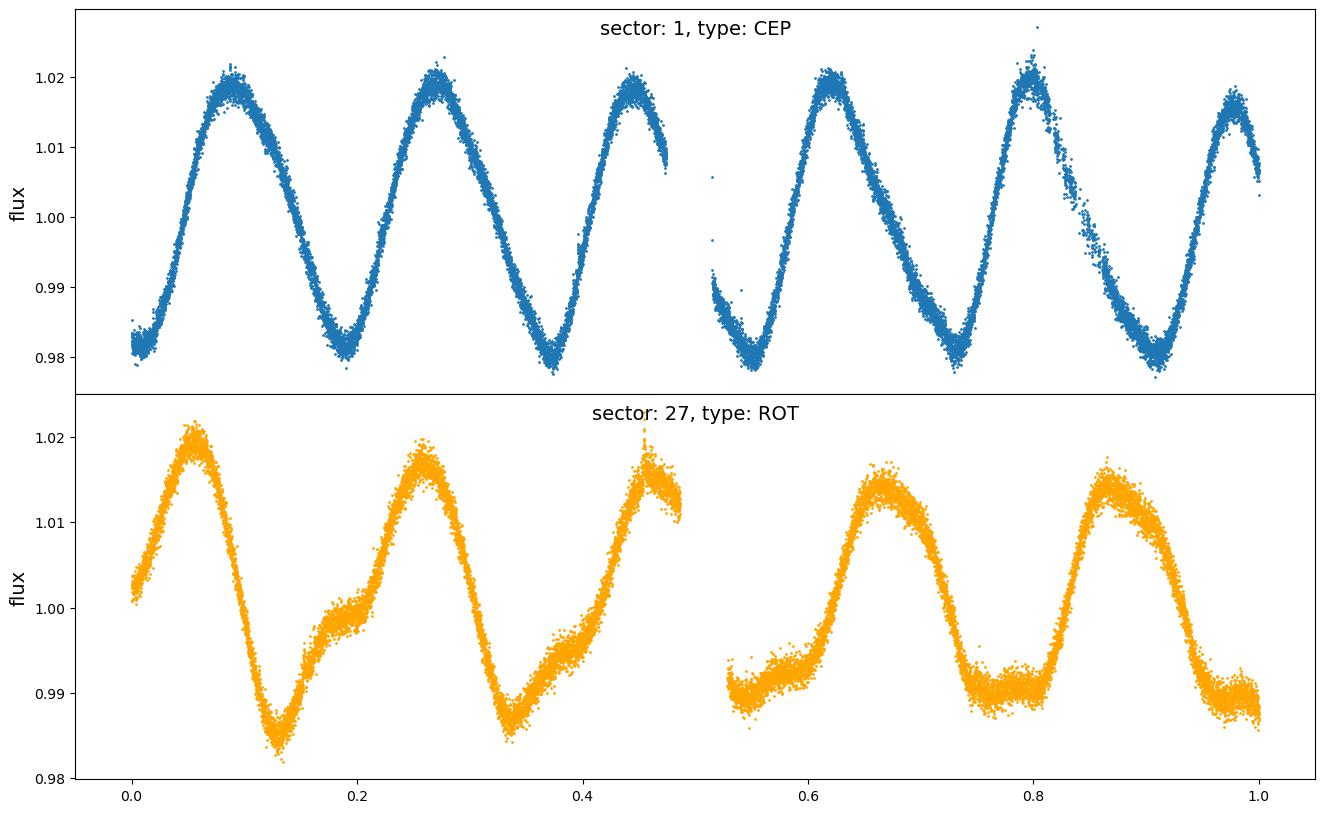}
    \caption{The figure shows the light curves of TIC 238853963 in sectors 1 and 27, exhibiting CEP and ROT variability, respectively. }
    \label{238853963}
\end{figure}

\begin{figure}
    \centering
    \includegraphics[width=1\linewidth]{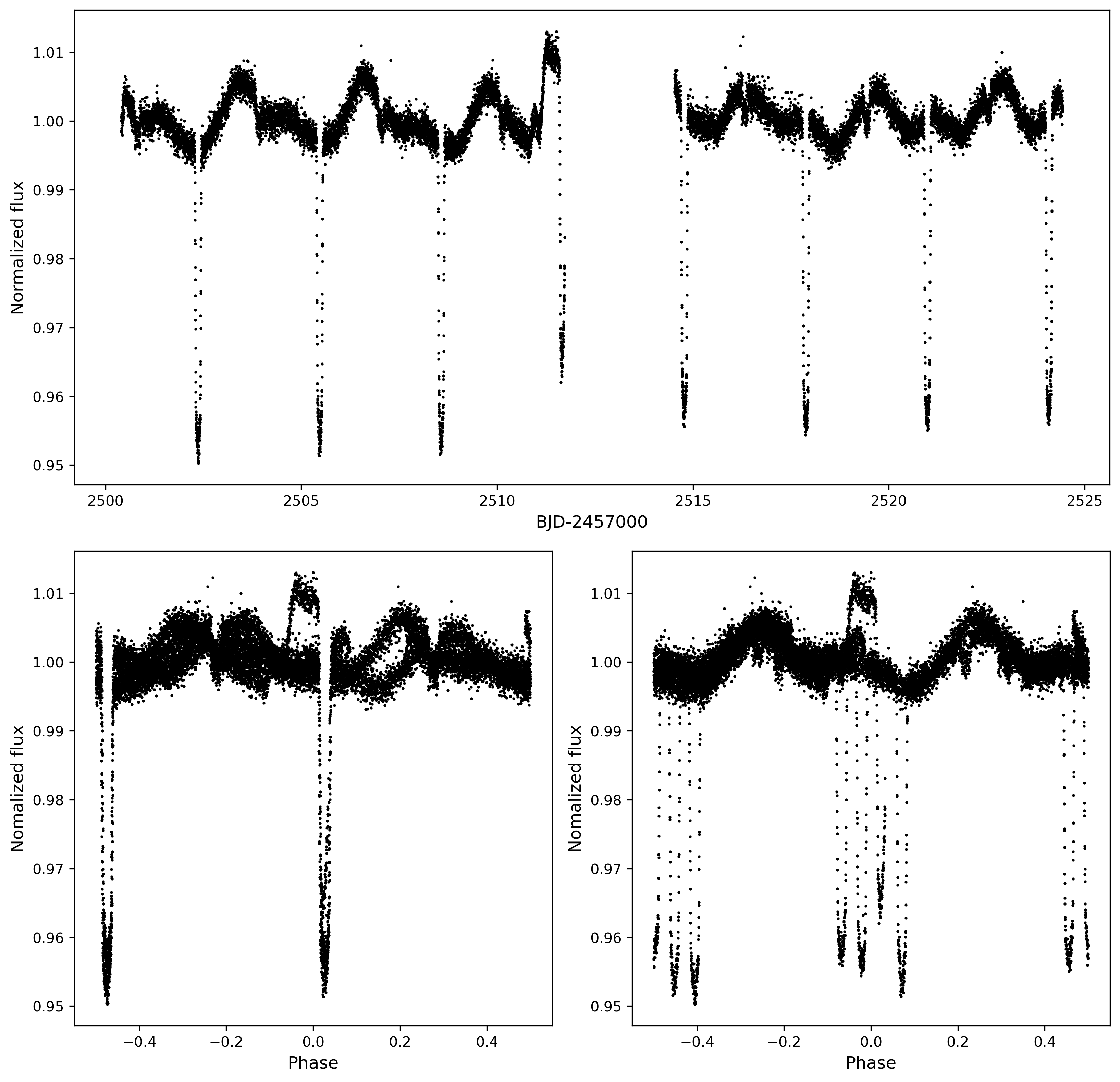}

    \caption{
The upper panel shows the original light curve, and the two lower panels display the phase-folded light curves at two different periods.  }
    \label{se_1_1}
\end{figure}

We summarized our classification results and conducted a visual inspection to further ensure the accuracy of the provided variable star catalog. We displayed typical curves of different types of variable stars in \Cref{fig:light}. In addition, We found that some targets exhibited different types of variability in different sectors, for example, \Cref{238853963} shows the different variability types of light curves of TIC 238853963 in two different sectors. 
We also find multi-periodic variables within a single sector, as shown in \Cref{se_1_1}. Clearly, the eclipsing period and the pulsation period are not consistent, indicating that the target has multiple periods. In this paper, we have chosen the most prominent period. 
Excluding these targets, we ultimately obtained a total of  6046 EA, 3859 EW, 2058 CEP, 8434 DSCT, 482 RRab, 416 RRc, and 9694 ROT. We provide two star catalogs: one containing targets that exhibit single variability, and the other comprising targets that show different variabilities of different TESS sectors, described as \Cref{dec} and \Cref{dec2}, respectively. The complete tables can be accessed via ChinaVO\footnote{\href{https://nadc.china-vo.org/res/r101519/}{DOI:10.12149/101519}}. By comparing the result of visual inspection with machine classification, we can identify some shortcomings in the classification process.


\begin{table}[h]
\caption{The variable star catalog containing targets that exhibit single variability.\footnote{(This table is available in its entirety in machine-readable form.)}}
\centering
\begin{tabular}{cc}
\hline
Name& Description\\
\hline
name&TESS Input Catalog ID\\
RF\_class& The classification results of the random forest\\
VI\_class& The classification results of the visual inspection\\
P&Period\\

 P\_c&The results of the visual inspection of periods, where correct is labeled as 1 and incorrect as 0.\\
 RA&Right Ascension\\
 DEC&Declination\\
 Amp&Amplitude\\
 Teff(K)&Temperature\\
 MAG&Magnitude\\
 sector&Sectors in which the target was observed\\
 $a_i$&The sine component of the $i$-th term in the Fourier series expansion\\
 $b_i$&The cosine component of the $i$-th term in the Fourier series expansion\\
 $r^2$&Goodness of fit\\
 \hline
\end{tabular}
  
\label{dec}
\end{table}

\begin{table}[h]
\caption{The variable star catalog containing targets that exhibit non-single variability.\footnote{(This table is available in its entirety in machine-readable form.)}}
\centering
\begin{tabular}{cc}
\hline
Name& Description\\
\hline
name&TESS Input Catalog ID\\
RF\_class& The classification results of the random forest\\
VI\_class& The classification results of the visual inspection\\
P&Period\\

 P\_c&The results of the visual inspection of periods, where correct is labeled as 1 and incorrect as 0\\
 RA&Right Ascension\\
 DEC&Declination\\
 Amp&Amplitude\\
 Teff(K)&Temperature\\
 MAG&Magnitude\\
 sector&Sectors in which the target was observed\\
 \hline
\end{tabular}

\label{dec2}
\end{table}

\begin{figure}
    \centering
    \includegraphics[width=1\linewidth]{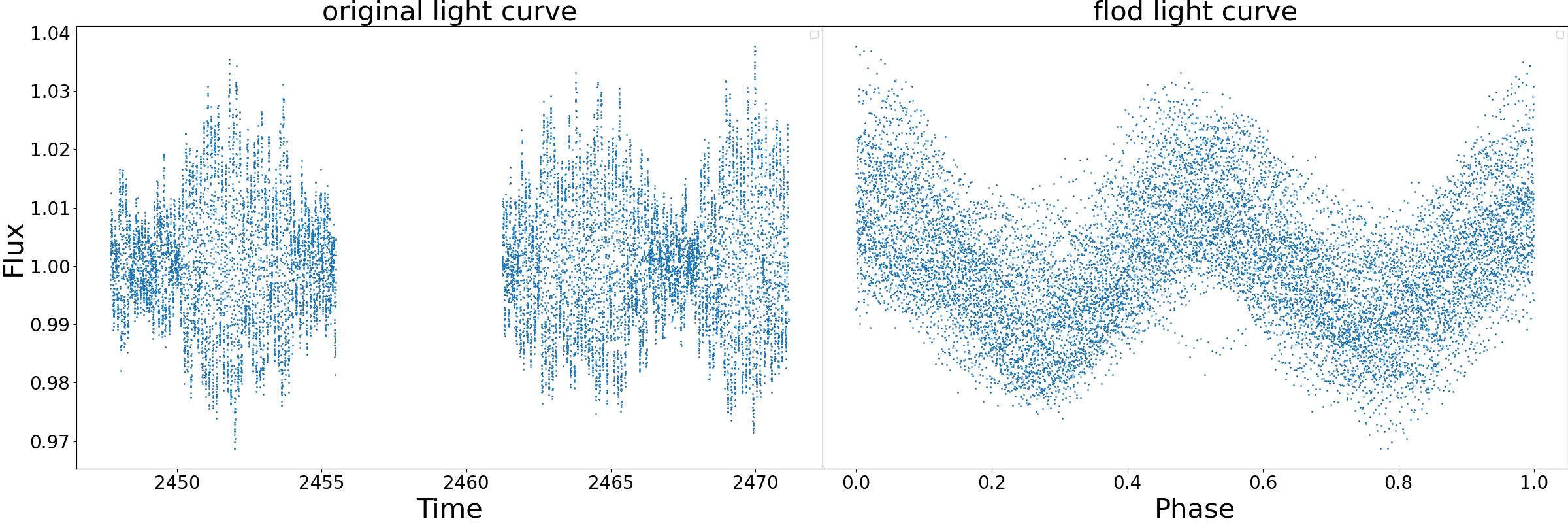}
    \caption{This light curve is a typical example of a CEP mixed within EW. The left panel shows the original light curve, while the right panel shows the phase folded light curve.}
    \label{fig:CEP_EW}
\end{figure}

A relatively high number of CEP stars were classified as EW. \Cref{fig:CEP_EW} shows a typical light curve of a CEP star. 
CEP stars display brightness variations resulting from non-radial surface pulsations. However, their phase folded light curves closely resemble those of EW stars, which significantly increases the difficulty of classification. Due to observational constraints, this type of variable star is frequently misclassified as EW by ground-based telescopes. 

Subsequently, due to the limited duration of the observation data, we limited the range of our period search. Therefore, some long-period variable stars, such as EA, inevitably exhibit period errors. In the variable star catalog we provided, we specifically annotated whether the period was accurate during the visual inspection process.  

\begin{figure}[ht]
    \centering
    \begin{minipage}[b]{0.49\textwidth}
        \centering
        \includegraphics[width=\textwidth]{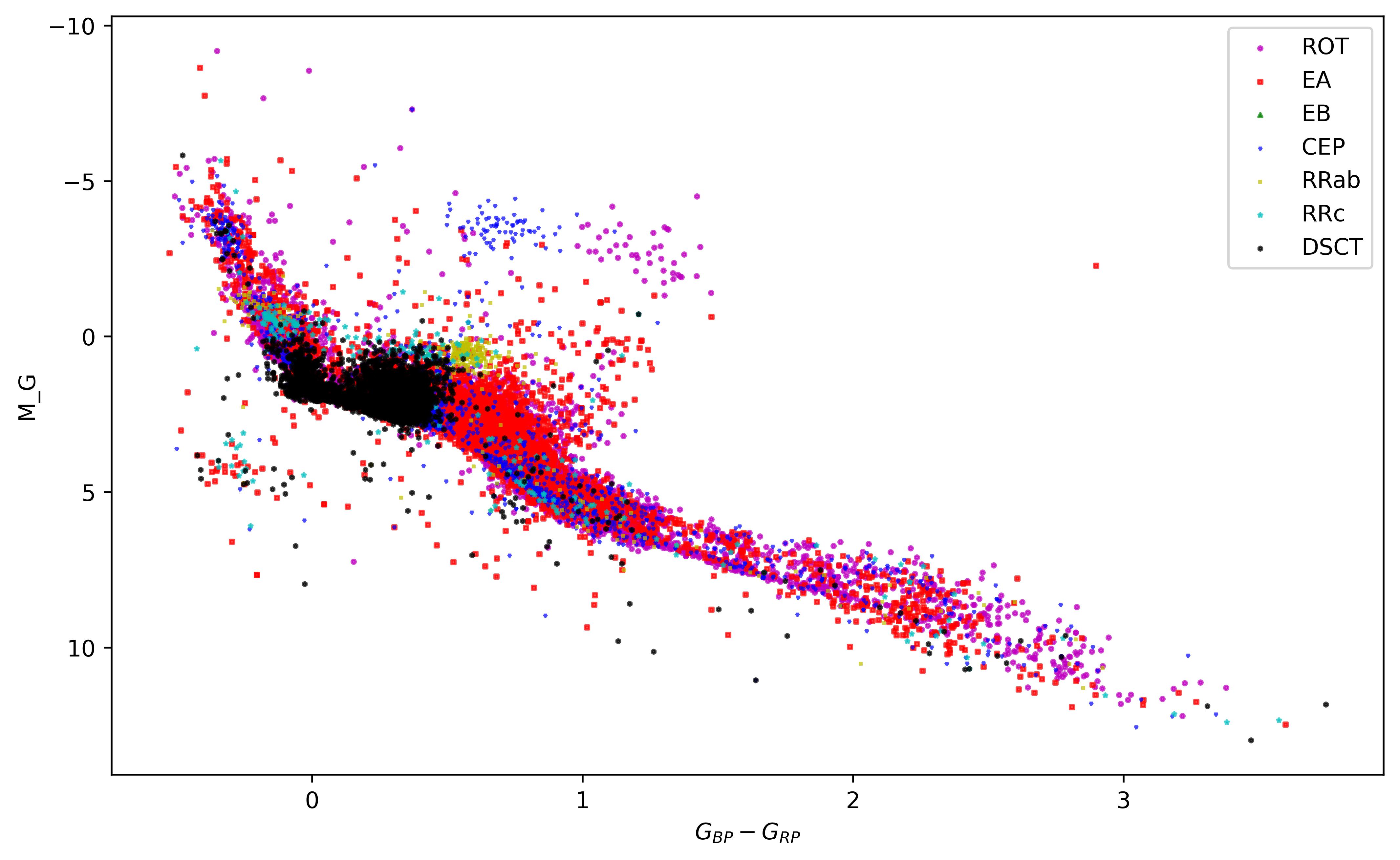}
    \end{minipage}
    \hfill
    \begin{minipage}[b]{0.49\textwidth}
        \centering
        \includegraphics[width=\textwidth]{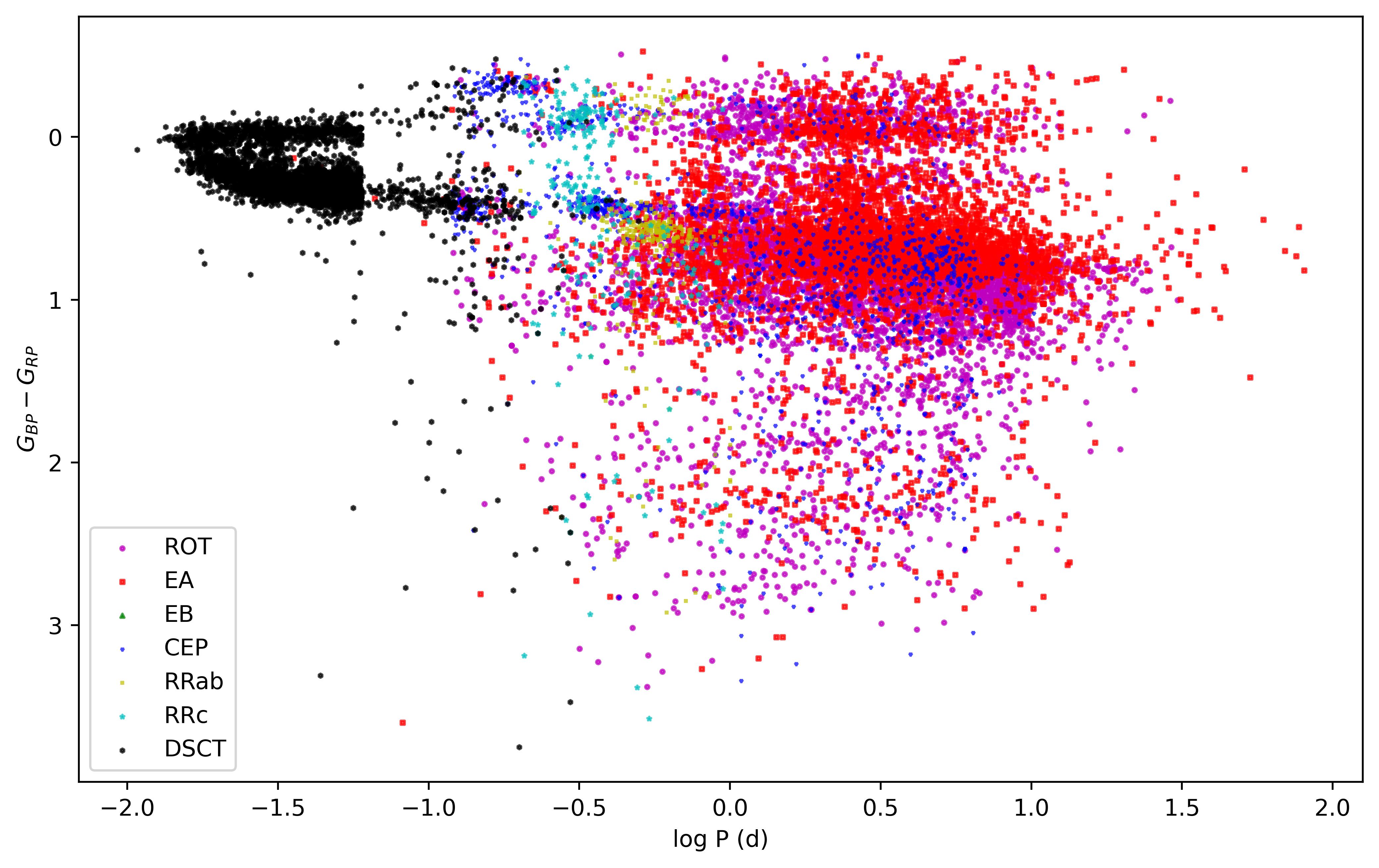}
    \end{minipage}
    \caption{The figure shows a H-R diagram (left) and color-period diagram (right) plotted using stars obtained from the intersection of our catalog with Gaia data.}
    \label{fig:HL}
\end{figure}

We cross-matched our variable star catalog with the Gaia DR3 catalog using a matching radius of 3 arcseconds and plotted the matched stars on the Hertzsprung-Russell (H-R) diagram and color-period diagram, as shown in \Cref{fig:HL}.

We compared the obtained $EB_s$ with the eclipsing binary catalog provided by \cite{EB_4000}, finding 3610 targets with consistent classifications, so we discovered 6245 new ones. We cross-matched the variable star catalog with the Gaia\citep{q1}, VSX \citep{vsx}, TESS \citep{EB_4000}, ZTF \citep{r8}, and ASAS-SN \citep{r2} variable catalogs, and a total of 16298 targets have been identified. Therefore, 14092 new variables are discovered.

\section{CONCLUSIONS} \label{sec:CONCLUSIONS}
In this paper, we applied machine learning method to classify variable stars using TESS 2-minute data. By cross-matching with  Gaia DR3 variable star catalog, we obtained labels for $EB_s$, pulsations, and ROT variable stars. We calculated the periods of all stars using the GLS method and derived fourth-order Fourier parameters and goodness of fit $r^2$ for each target through Fourier fitting. We preliminarily selected non-variable stars based on power and goodness-of-fit. 
We trained a RF classifier using the labeled data and conducted an initial rough classification on the entire dataset. 
However, due to structural differences between the training data and the real data, we designed specific features for each type of light curve and conducted sub-class classifications.

After completion of all classifications, we performed a visual inspection of all variables to assess the accuracy of their classifications and periods while updating the provided catalog with relevant annotations. Finally, we provides a catalog of variable stars, comprising 6046 EA, 3859 EW, 2058 CEP, 8434 DSCT, 482 RRab, 416 RRc, and 9694 ROT, and a total of 14092 new variables are discovered.

\section*{ACKNOWLEDGEMENTS}
We sincerely thank the anonymous reviewer for the insightful comments and constructive suggestions, which have substantially improved.
This work was supported by National Natural Science Foundation of China (NSFC) (No. 12273018), and the Joint Research Fund in Astronomy (No. U1931103) under cooperative agreement between NSFC and Chinese Academy of Sciences (CAS), and by the Qilu Young Researcher Project of Shandong University, and by Young Data Scientist Program of the China National Astronomical Data Center Center and by the Cultivation Project for LAMOST Scientific Payoff and Research Achievement of CAMS-CAS. The calculations in this work were carried out at Supercomputing Center of Shandong University, Weihai.

This work includes data collected by the TESS mission. Funding for the TESS mission is provided by NASA Science Mission Directorate. We acknowledge the TESS team for its support of this work.

This research has made use of the VizieR catalogue access tool, CDS, Strasbourg, France. This research also made use of Astropy, a community-developed core Python package for Astronomy (Astropy Collaboration, 2013).

\appendix

\setcounter{figure}{0}
\renewcommand{\thefigure}{A\arabic{figure}}

\begin{figure}[h]
    \centering
    \includegraphics[width=1\linewidth]{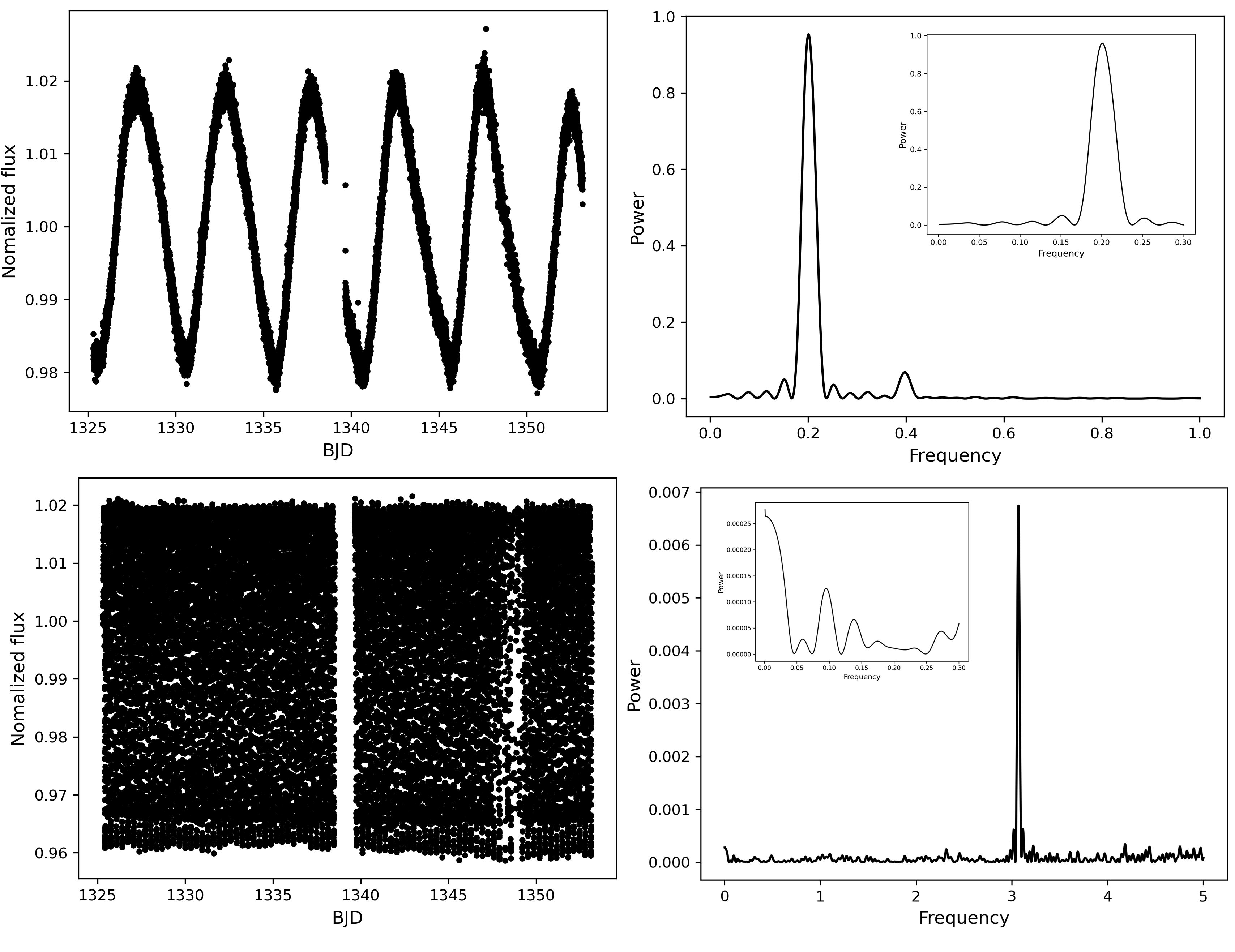}
    \caption{The left panel shows the light curve, and the right panel shows the GLS periodogram and the inset is an enlargement of the area around the 7 days period. }
    \label{fig:LS_p}
\end{figure}

\begin{figure}
    \centering
    \includegraphics[width=1\linewidth]{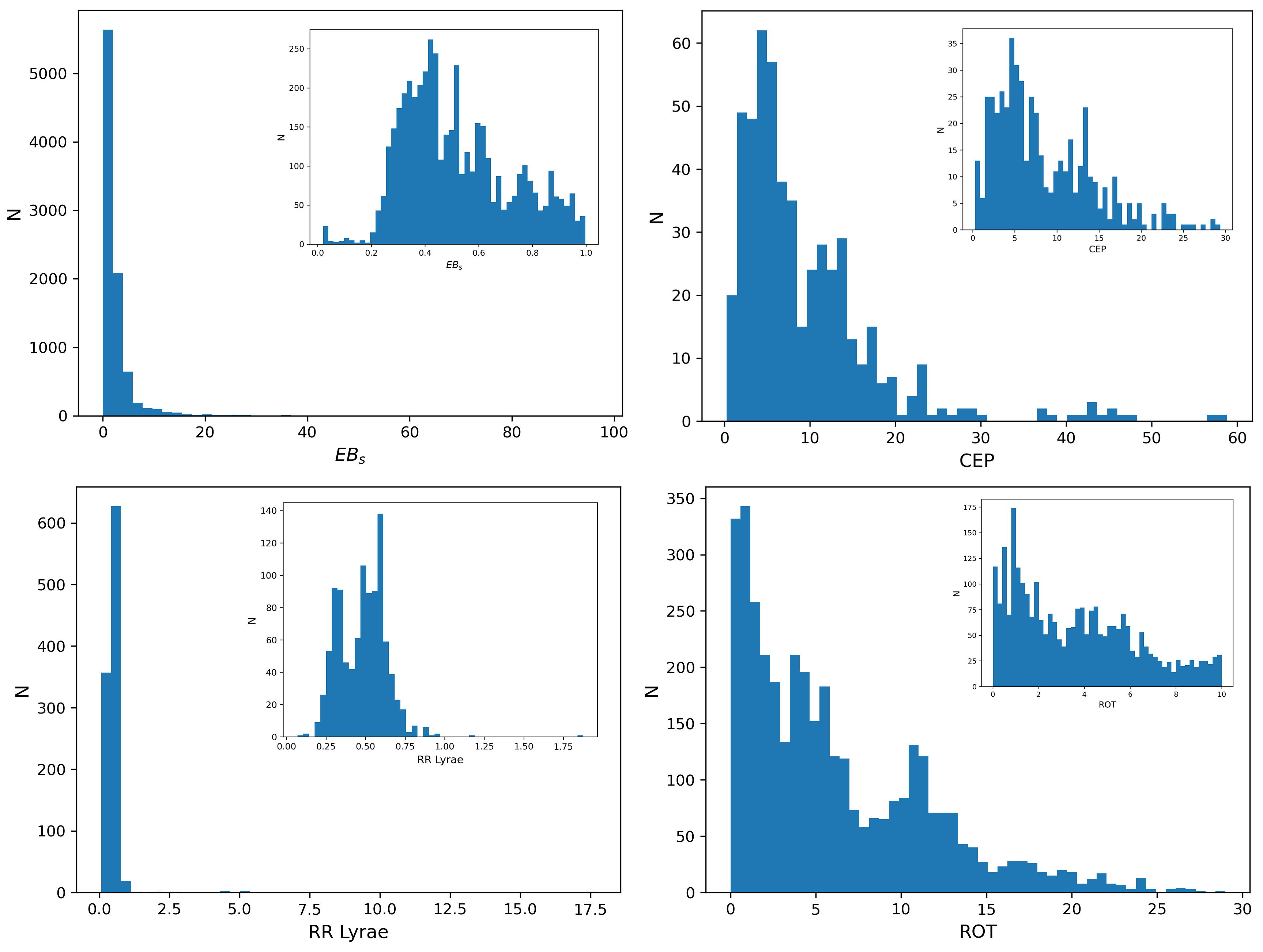}
    \caption{\textbf{
}The distribution of periods for four different categories of variable stars. }
    \label{fig:p_dis}
\end{figure}

\begin{figure}
    \centering
    \includegraphics[width=0.5\linewidth]{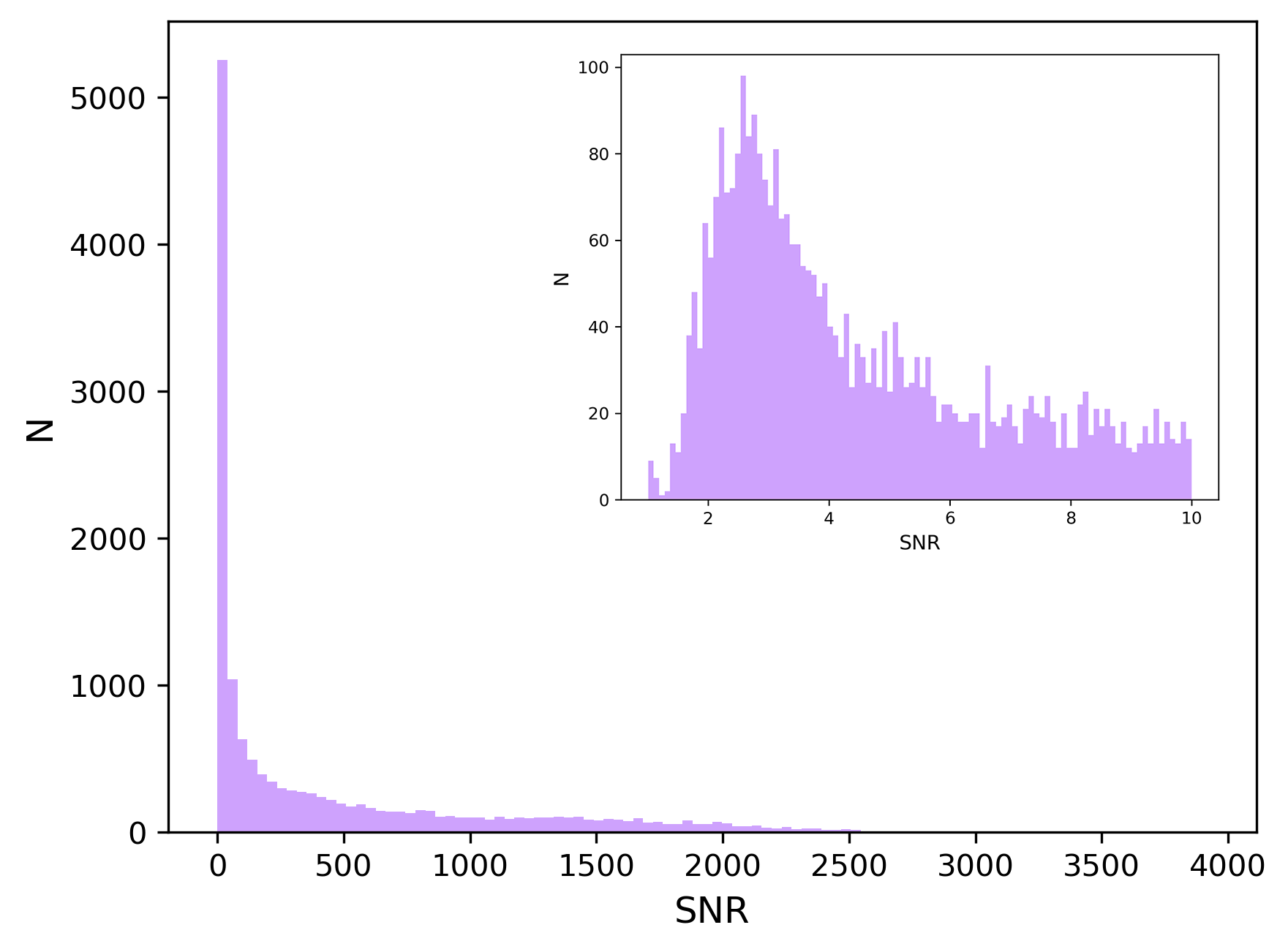}
    \caption{\textbf{
}The SNR distribution was obtained through LS periodograms.}
    \label{fig:snr}
\end{figure}

\bibliography{sample631}{}
\bibliographystyle{aasjournal}



\end{document}